\documentstyle[aps,prd,epsfig]{revtex}

\begin{document}
\draft

\title{Longitudinal gluons and Nambu-Goldstone bosons 
in a two-flavor color superconductor}

\author{Dirk H. Rischke}
\address{Institut f\"ur Theoretische Physik, 
Johann Wolfgang Goethe-Universit\"at \\
Robert-Mayer-Str.\ 8--10, D-60054 Frankfurt/Main, Germany \\
E-mail: drischke@th.physik.uni-frankfurt.de}

\author{Igor A. Shovkovy\thanks{On leave of absence from 
Bogolyubov Institute for Theoretical Physics, 252143 Kiev, Ukraine.}}
\address{School of Physics and Astronomy,
University of Minnesota \\
116 Church Street S.E., Minneapolis, MN 55455, U.S.A. \\
E-mail: shovkovy@physics.umn.edu}

\date{\today}
\maketitle

\begin{abstract} 
In a two-flavor color superconductor, the $SU(3)_c$ gauge symmetry
is spontaneously broken by diquark condensation.
The Nambu-Goldstone excitations of the diquark condensate 
mix with the gluons associated with
the broken generators of the original gauge group.
It is shown how one can decouple these modes with 
a particular choice of 't~Hooft gauge.
We then explicitly compute the spectral density
for transverse and longitudinal gluons of adjoint color 8.
The Nambu-Goldstone excitations give rise to a
singularity in the real part of the longitudinal gluon self-energy.
This leads to a vanishing gluon spectral density for energies
and momenta located on the dispersion branch of the 
Nambu-Goldstone excitations.
\end{abstract}

\section{Introduction}

Cold, dense quark matter is a color superconductor \cite{bailinlove}.
For two massless quark flavors (say, up and down), Cooper pairs 
with total spin zero condense in the color-antitriplet, 
flavor-singlet channel. In this so-called two-flavor color superconductor,
the $SU(3)_c$ gauge symmetry is spontaneously broken to 
$SU(2)_c$ \cite{arw}. If we choose to orient the (anti-) color
charge of the Cooper pair along the (anti-) blue direction
in color space, only red and green quarks form
Cooper pairs, while blue quarks remain unpaired. Then,
the three generators $T_1,\, T_2,$ and $T_3$ of the
original $SU(3)_c$ gauge group form the generators of the
residual $SU(2)_c$ symmetry. The remaining five generators 
$T_4, \ldots, T_8$ are broken.
(More precisely, the last broken generator is a combination of $T_8$ and 
the generator ${\bf 1}$ of the global $U(1)$ symmetry of baryon number
conservation, for details see Ref.\ \cite{sw2} and below).

According to Goldstone's theorem, this pattern of 
symmetry breaking gives rise to five massless bosons, the so-called
Nambu-Goldstone bosons, corresponding to the five broken 
generators of $SU(3)_c$. Physically, these massless bosons correspond
to fluctuations of the order parameter, in our case the diquark
condensate, in directions in color-flavor space where
the effective potential is flat.
For gauge theories (where the local gauge symmetry cannot truly
be spontaneously broken), these bosons 
are ``eaten'' by the gauge bosons corresponding to the broken
generators of the original gauge group, {\it i.e.}, in our case the
gluons with adjoint colors $a= 4, \ldots, 8$. They give rise
to a longitudinal degree of freedom for these gauge bosons.
The appearance of a longitudinal degree of freedom is
commonly a sign that the gauge boson becomes massive.

In a dense (or hot) medium, however, even {\em without\/} spontaneous
breaking of the gauge symmetry the gauge bosons already
have a longitudinal degree of freedom, the so-called 
{\em plasmon\/} mode \cite{LeBellac}. Its appearance is related
to the presence of gapless charged quasiparticles.
Both transverse and longitudinal modes exhibit a mass gap,
{\it i.e.}, the gluon energy $p_0 \rightarrow m_g > 0$ for
momenta $p \rightarrow 0$.
In quark matter with $N_f$ massless quark flavors
at zero temperature $T=0$, the gluon mass parameter 
(squared) is \cite{LeBellac}
\begin{equation} \label{gluonmass}
m_g^2 = \frac{N_f}{6\, \pi^2} \, g^2 \, \mu^2\,\, ,
\end{equation}
where $g$ is the QCD coupling constant and 
$\mu$ is the quark chemical potential.

It is {\em a priori\/} unclear
how the Nambu-Goldstone bosons interact with these
longitudinal gluon modes. In particular, it is of
interest to know whether coupling 
terms between these modes exist and, if yes, whether
these terms can be eliminated by a suitable choice of ('t~Hooft) gauge.
The aim of the present work is to address these questions.
We shall show that the answer to both questions is ``yes''.
We shall then demonstrate by focussing on the gluon of adjoint color 8,
how the Nambu-Goldstone mode affects the spectral density
of the longitudinal gluon.

Our work is partially based on and motivated by previous 
studies of gluons in a two-flavor color superconductor
\cite{carterdiakonov,dhr2f,dhrselfenergy}.
The gluon self-energy and the resulting spectral properties 
have been discussed in Ref.\ \cite{dhrselfenergy}. In that paper, however,
the fluctuations of the diquark condensate have been neglected.
Consequently, the longitudinal degrees of freedom of the gluons
corresponding to the broken generators of $SU(3)_c$ have not been
treated correctly. The gluon polarization tensor was no longer
explicitly transverse (a transverse polarization tensor $\Pi^{\mu\nu}$
obeys $P_\mu \, \Pi^{\mu \nu} = \Pi^{\mu \nu}\, P_\nu = 0$),
and it did not satisfy the Slavnov-Taylor identity. 
As a consequence, the plasmon mode exhibited a certain peculiar
behavior in the low-momentum limit, which cannot be physical
(cf.\ Fig.\ 5 (a) of Ref.\ \cite{dhrselfenergy}).
It was already realized in Ref.\ \cite{dhrselfenergy}
that the reason for this unphysical behavior is the
fact that the mixing of the gluon with the excitations
of the condensate was neglected. It was moreover suggested 
in Ref.\ \cite{dhrselfenergy} that proper inclusion of this mixing would
amend the shortcomings of the previous analysis.
The aim of the present work is to follow this suggestion and
thus to correct the results of Ref.\ \cite{dhrselfenergy} with respect 
to the longitudinal gluon. Note that in Ref.\ \cite{carterdiakonov}
fluctuations of the color-superconducting condensate were 
taken into account in the calculation of the gluon polarization
tensor. As a consequence, the latter is explicitly transverse.
However, the analysis was done in the vacuum, at $\mu=0$, not at
(asymptotically) large chemical potential.

The outline of the present work is as follows.
In Section \ref{II} we derive the transverse and
longitudinal gluon propagators 
including fluctuations of the diquark condensate.
In Section \ref{III} we use the resulting expressions to
compute the spectral density
for the gluon of adjoint color 8. Section \ref{IV}
concludes this work with a summary of our results.

Our units are $\hbar=c=k_B=1$. The metric tensor is
$g^{\mu \nu}= {\rm diag}\,(+,-,-,-)$. We denote 4-vectors in
energy-momentum space by capital letters,
$K^{\mu} = (k_0,{\bf k})$. Absolute magnitudes of
3-vectors are denoted as $k \equiv |{\bf k}|$, and the unit
vector in the direction of ${\bf k}$ is $\hat{\bf k} \equiv
{\bf k}/k$.

\section{Derivation of the propagator for transverse and longitudinal gluons}
\label{II}

In this section, we derive the gluon propagator taking
into account the fluctuations of the diquark condensate.
A short version of this derivation can be found in 
Appendix C of Ref.\ \cite{msw} [see also the original Ref.\ \cite{gusyshov}].
Nevertheless, for the sake of clarity and in order to make our presentation
self-contained, we decide to present this once more in greater detail
and in the notation of Ref.\ \cite{dhrselfenergy}.
As this part is rather technical, the reader less interested in the
details of the derivation should skip directly to our main result,
Eqs.\ (\ref{transverse}), (\ref{longitudinal}), and (\ref{hatPi00aa}).

We start with the grand partition function of QCD,
\begin{mathletters} \label{Z}
\begin{equation} \label{ZQCD}
{\cal Z} = \int {\cal D} A \; e^{ S_A }
\;{\cal Z}_q[A]\,\, ,  
\end{equation}
where
\begin{equation}
{\cal Z}_q[A] =  \int {\cal D} \bar{\psi} \, {\cal D} \psi\,
\exp \left[ \int_x  \bar{\psi} \left(
i \gamma^\mu \partial_\mu + \mu \gamma_0 + g \gamma^\mu A_\mu^a T_a
\right) \psi \right]
\,. \label{Zquarks}
\end{equation}
\end{mathletters}
is the grand partition function for massless quarks in the presence of
a gluon field $A^\mu_a$. In Eq.\ (\ref{Z}), the
space-time integration is defined as
$\int_x \equiv \int_0^{1/T} d\tau \int_V d^3{\bf x}\,$,
where $V$ is the volume of the system, 
$\gamma^\mu$ are the Dirac matrices, and $T_a= \lambda_a/2$ are
the generators of $SU(N_c)$. For QCD, $N_c = 3$, and $\lambda_a$ are the
Gell-Mann matrices. The quark fields $\psi$ 
are $4 N_c N_f$-component spinors, {\it i.e.}, 
they carry Dirac indices $\alpha = 1, \ldots,4$, fundamental color indices
$i=1,\ldots,N_c$, and flavor indices $f=1,\ldots,N_f$.
The action for the gauge fields consists of three parts,
\begin{equation} \label{L_A}
S_A = S_{F^2} + S_{\rm gf} + S_{\rm FPG}\,\, ,
\end{equation}
where 
\begin{equation}
S_{F^2} = - \frac{1}{4} \int_x F^{\mu \nu}_a \, F_{\mu \nu}^a
\end{equation}
is the gauge field part; here, $F_{\mu\nu}^a = \partial_\mu A_\nu^a 
- \partial_\nu A_\mu^a + g f^{abc} A_\mu^b A_\nu^c$ is the
field strength tensor. The part corresponding to gauge fixing,
$S_{\rm gf}$, and to Fadeev-Popov ghosts, $S_{\rm FPG}$,
will be discussed later.

For fermions at finite chemical potential it is advantageous
to introduce the charge-conjugate degrees of freedom explicitly.
This restores the symmetry of the theory under $\mu \rightarrow - \mu$.
Therefore, in Ref.\ \cite{dhr2f}, a kind of replica method was applied,
in which one first artificially increases 
the number of quark species, and then replaces half of these
species of quark fields by charge-conjugate quark fields.
More precisely, first replace the quark partition function 
${\cal Z}_q[A]$ by
${\cal Z}_M[A] \equiv \left\{ {\cal Z}_q[A] \right\}^M$, 
$M$ being some large integer number. (Sending $M \rightarrow 1$
at the end of the calculation reproduces the original
partition function.)
Then, take $M$ to be an even integer number, and replace 
the quark fields by charge-conjugate quark fields in $M/2$ of
the factors ${\cal Z}_q[A]$ in ${\cal Z}_M[A]$. This results in
\begin{equation}
{\cal Z}_M[A]  =  
\int \prod_{r=1}^{M/2} {\cal D} \bar{\Psi}_r \, {\cal D} \Psi_r \; \exp 
\left\{ \sum_{r=1}^{M/2} \left[
\int_{x,y} \bar{\Psi}_r(x) \,{\cal G}_0^{-1} (x,y)\,
\Psi_r(y) + \int_x  g\, \bar{\Psi}_r(x) \,  A_\mu^a(x)\,\hat{\Gamma}^\mu_a\,
\Psi_r(x) \right] \right\} \,\, . \label{Zquarks2}
\end{equation}
Here, $r$ labels the quark species and 
$\Psi_r$, $\bar{\Psi}_r$ are $8 N_c N_f$-component Nambu-Gor'kov spinors,
\begin{equation}
\Psi_r \equiv \left( \begin{array}{c} 
                    \psi_r \\
                    \psi_{C r} 
                   \end{array}
            \right) \,\,\, , \,\,\,\,
\bar{\Psi}_r \equiv ( \bar{\psi}_r \, , \, \bar{\psi}_{C r} )\,\, , 
\end{equation}
where $\psi_{C r} \equiv C \bar{\psi}_r^T$ is the charge
conjugate spinor and $C=i \gamma^2 \gamma_0$ is the charge conjugation
matrix. The inverse of the $8 N_c N_f \times 8 N_c N_f$-dimensional
Nambu-Gor'kov propagator for non-interacting quarks is defined as
\begin{equation} \label{S0-1}
{\cal G}_0^{-1} \equiv
\left( \begin{array}{cc}
            [G_0^+]^{-1} & 0 \\
             0 & [G_0^-]^{-1}
       \end{array} \right)\,\, ,
\end{equation}
where
\begin{equation} \label{G0pm-1}
[G_0^\pm]^{-1}(x,y) \equiv -i \left( i \gamma_\mu \partial^\mu_x
\pm \mu \gamma_0 \right) \delta^{(4)}(x-y) 
\end{equation}
is the inverse propagator for non-interacting quarks (upper sign) or 
charge conjugate quarks (lower sign), respectively.
The Nambu-Gor'kov matrix vertex describing the interaction between 
quarks and gauge fields is defined as follows:
\begin{equation} \label{Gamma}
\hat{\Gamma}^\mu_a \equiv \left( \begin{array}{cc}
                                 \Gamma^\mu_a & 0 \\
                                 0 & \bar{\Gamma}^\mu_a
                          \end{array} \right)\,\, ,
\end{equation}
where $\Gamma^\mu_a \equiv \gamma^\mu T_a$ and
$\bar{\Gamma}^\mu_a \equiv C (\gamma^\mu)^T C^{-1} T_a^T \equiv -\gamma^\mu
T_a^T$.

Following Ref.\ \cite{bailinlove} we now add the term
$\int_{x,y} \bar{\psi}_{C r}(x)\, \Delta^+(x,y) \,
\psi_r(y)$ and the corresponding charge-conjugate term
$\int_{x,y} \bar{\psi}_r(x)\, \Delta^-(x,y) \, \psi_{C r}(y)$,
where $\Delta^- \equiv \gamma_0 \, (\Delta^+)^\dagger \, \gamma_0$,
to the argument of the exponent in Eq.\ (\ref{Zquarks2}).
This defines the quark (replica) partition function in the presence of
the gluon field $A^\mu_a$ {\it and\/} the diquark source
fields $\Delta^+$, $\Delta^-$:
\begin{equation}
{\cal Z}_M[A,\Delta^+,\Delta^-] \equiv
\int \prod_{r=1}^{M/2} {\cal D} \bar{\Psi}_r \, {\cal D} \Psi_r \; \exp 
\left\{ \sum_{r=1}^{M/2} \left[
\int_{x,y} \bar{\Psi}_r(x) \,{\cal G}^{-1} (x,y)\,
\Psi_r(y) + \int_x  g\, \bar{\Psi}_r(x) \,  A_\mu^a(x)\,\hat{\Gamma}^\mu_a\,
\Psi_r(x) \right] \right\} \,\, , \label{Zquarks3}
\end{equation}
where
\begin{equation}\label{G-1}
{\cal G}^{-1} \equiv
\left( \begin{array}{cc}
            [G^+_0]^{-1} & \Delta^- \\
             \Delta^+ & [G^-_0]^{-1}
       \end{array} \right)
\end{equation}
is the inverse quasiparticle propagator.

Inserting the partition function (\ref{Zquarks3}) into
Eq.\ (\ref{ZQCD}), the (replica) QCD partition function is then computed in
the presence of the (external) diquark source terms $\Delta^\pm(x,y)$,
${\cal Z} \rightarrow {\cal Z}[\Delta^+,\Delta^-]$.
In principle, this is not the physically relevant quantity, from which
one derives thermodynamic properties of the color superconductor.
The diquark condensate is not an external field, but
assumes a nonzero value because of an intrinsic property of the system,
namely the attractive gluon interaction in the color-antitriplet
channel, which destabilizes the Fermi surface. 

The proper functional from which one derives thermodynamic 
functions is obtained by a Legendre transformation of  
$\ln {\cal Z} [\Delta^+, \Delta^-]$, in which the functional dependence on 
the diquark source term is replaced by that on the corresponding canonically
conjugate variable, the diquark condensate. The Legendre-transformed
functional is the effective action for the diquark condensate. 
If the latter is {\em constant}, the effective action is,
up to a factor of $V/T$, identical to the effective potential.
The effective potential is simply a function of the diquark condensate.
Its explicit form for large-density QCD was derived in Ref.\ 
\cite{eff-pot}. The value of this function at its maximum determines the
pressure. The maximum is determined by a Dyson-Schwinger
equation for the diquark condensate, which is identical to
the standard gap equation for the color-superconducting gap. 
It has been solved in the mean-field approximation in Refs.\ 
\cite{rdpdhr2,schaferwilczek,miransky}.
In the mean-field approximation \cite{rdpdhrscalar},
\begin{equation} \label{mfa}
\Delta^+(x,y) \sim
\langle \, \psi_{C r}(x) \, \bar{\psi}_r(y)\, \rangle 
\,\,\,\, , \,\,\,\,\,
\Delta^-(x,y) \sim
\langle \, \psi_r(x) \, \bar{\psi}_{C r}(y) \, \rangle \,\, .
\end{equation}
In this work, we are interested in the gluon propagator,
and the derivation of the pressure via a Legendre transformation
of $\ln {\cal Z}[\Delta^+,\Delta^-]$ is of no concern to us. 
In the following, we shall therefore 
continue to consider the partition function in the presence of 
(external) diquark source terms $\Delta^\pm$.

The diquark source terms in the quark (replica) partition function
(\ref{Zquarks3}) could in principle be chosen
differently for each quark species. This could be made explicit by
giving $\Delta^\pm$ a subscript $r$, $\Delta^\pm \rightarrow 
\Delta^\pm_r$. However, as we take the limit $M \rightarrow 1$
at the end, it is not necessary to do so, as only $\Delta^\pm_1
\equiv \Delta^\pm$ will survive anyway. In other words, we 
use the {\em same\/} diquark sources for {\em all\/} quark species.

The next step is to explicitly investigate the
fluctuations of the diquark condensate around its
expectation value. These fluctuations correspond physically to
the Nambu-Goldstone excitations (loosely termed ``mesons'' in the
following) in a color superconductor. As mentioned in the
introduction, there are five such mesons
in a two-flavor color superconductor, 
corresponding to the generators of $SU(3)_c$ which are broken
in the color-superconducting phase. If the condensate
is chosen to point in the (anti-) blue direction in fundamental
color space, the broken generators are $T_4, \ldots, T_7$
of the original $SU(3)_c$ group and the particular
combination $B \equiv ({\bf 1} + \sqrt{3} T_8)/3$ of generators of
the global $U(1)_B$ and local $SU(3)_c$ symmetry \cite{sw2}.

The effective action for the diquark condensate and, consequently,
for the meson fields as fluctuations of the diquark condensate, is
derived via a Legendre transformation of $\ln {\cal Z}[\Delta^+,\Delta^-]$.
In this work, we are concerned with the properties of the gluons and
thus refrain from computing this effective action explicitly.
Consequently, instead of considering the physical meson fields,
we consider the variables in ${\cal Z}[\Delta^+,\Delta^-]$, which correspond 
to these fields. These are the fluctuations of the diquark 
source terms $\Delta^\pm$. 
We choose these fluctuations to be
complex phase factors multiplying the magnitude of the source terms,
\begin{mathletters} \label{DeltaPhi}
\begin{eqnarray}
\Delta^+(x,y) & = & 
{\cal V}^* (x)\, \Phi^+(x,y) \, {\cal V}^\dagger(y) \,\, , \\
\Delta^-(x,y) & = & 
{\cal V}(x) \, \Phi^-(x,y) \, {\cal V}^T(y) \,\, ,
\end{eqnarray}
\end{mathletters}
where
\begin{equation} \label{phase}
{\cal V}(x) \equiv \exp \left[ i \left( \sum_{a=4}^7 \varphi_a(x) T_a
+ \frac{1}{\sqrt{3}}\, \varphi_8(x) B \right) \right]\,\,.
\end{equation}
The extra factor $1/\sqrt{3}$ in front of $\varphi_8$ 
as compared to the treatment in Ref.\ \cite{msw} is chosen
to simplify the notation in the following.

Although the fields $\varphi_a$ are not the
meson fields themselves, but external fields which,
after a Legendre transformation of $\ln {\cal Z}[\Delta^+,\Delta^-]$, 
are replaced by the meson fields, we nevertheless 
(and somewhat imprecisely) refer to them as meson fields
in the following.
After having explicitly introduced the fluctuations of the
diquark source terms in terms of phase factors,
the functions $\Phi^\pm$ are only allowed to fluctuate in magnitude.
For the sake of completeness, let us mention that one could again
have introduced different fields 
$\varphi_{a r}$ for each replica $r$, but this is not really
necessary, as we shall take the limit $M \rightarrow 1$
at the end of the calculation anyway.

It is advantageous to also subject the quark fields $\psi_r$ to a
nonlinear transformation, introducing new fields $\chi_r$ via
\begin{equation} \label{chi}
\psi_r = {\cal V}\, \chi_r 
\,\,\,\, , \,\,\,\,\,
\bar{\psi}_r = \bar{\chi}_r\, {\cal V}^\dagger\,\, .
\end{equation}
Since the meson fields are real-valued and the
generators $T_4, \ldots, T_7$ and $B$ are hermitian, the (matrix-valued) 
operator ${\cal V}$ is unitary, ${\cal V}^{-1} = {\cal V}^\dagger$.
Therefore, the measure of the Grassmann integration over quark fields
in Eq.\ (\ref{Zquarks3}) remains unchanged. From Eq.\ (\ref{chi}), 
the charge-conjugate fields transform as
\begin{equation}
\psi_{C r} = {\cal V}^* \, \chi_{C r}
\,\,\,\, , \,\,\,\,\,
\bar{\psi}_{C r} = \bar{\chi}_{C r} \, {\cal V}^T\,\, ,
\end{equation}

The advantage of transforming the quark fields is
that this preserves the simple structure of the terms coupling
the quark fields to the diquark sources,
\begin{equation}
\bar{\psi}_{C r}(x)\, \Delta^+(x,y) \, \psi_r(y)
\equiv \bar{\chi}_{C r}(x)\, \Phi^+(x,y) \, \chi_r(y)
\,\,\,\, , \,\,\,\,\,
\bar{\psi}_r(x)\, \Delta^-(x,y) \, \psi_{C r}(y)
\equiv \bar{\chi}_r(x)\, \Phi^-(x,y) \, \chi_{C r}(y) \,\, .
\end{equation}
In mean-field approximation, the diquark source
terms are proportional to
\begin{equation} \label{mfa2}
\Phi^+(x,y) 
\sim \langle \, \chi_{C r}(x) \, \bar{\chi}_r(y)\, \rangle 
\,\,\,\, , \,\,\,\,\,
\Phi^-(x,y) 
\sim \langle \, \chi_r(x) \, \bar{\chi}_{C r}(y) \, \rangle\,\, .
\end{equation}

The transformation (\ref{chi}) has the following effect on
the kinetic terms of the quarks and the term coupling quarks to gluons:
\begin{mathletters}
\begin{eqnarray}
\bar{\psi}_r \, \left( i \, \gamma^\mu \partial_\mu
+ \mu \, \gamma_0 + g \, \gamma_\mu \, A^\mu_a T_a \right)\,
\psi_r & = & \bar{\chi}_r \, \left( i\, \gamma^\mu \partial_\mu
+ \mu \, \gamma_0 + \gamma_\mu \, \omega^\mu \right) \, \chi_r \,\, , \\
\bar{\psi}_{C r} \, \left( i \, \gamma^\mu \partial_\mu
- \mu \, \gamma_0 - g \, \gamma_\mu \, A^\mu_a T_a^T \right)\,
\psi_{C r} & = & \bar{\chi}_{C r} \, \left( i\, \gamma^\mu \partial_\mu
- \mu \, \gamma_0 + \gamma_\mu \, \omega^\mu_C \right) \, \chi_{C r} \,\, ,
\end{eqnarray}
\end{mathletters}
where
\begin{mathletters} \label{Maurer1}
\begin{equation}
\omega^\mu \equiv {\cal V}^\dagger \, \left( i \, \partial^\mu
+ g\, A^\mu_a T_a \right) \, {\cal V}
\end{equation}
is the $N_c N_f \times N_c N_f$-dimensional
Maurer-Cartan one-form introduced in Ref.\ \cite{sannino} and
\begin{equation} 
\omega^\mu_C \equiv {\cal V}^T \, \left( i \, \partial^\mu
- g\, A^\mu_a T_a^T \right) \, {\cal V}^* 
\end{equation}
\end{mathletters}
is its charge-conjugate version. Note that the partial derivative
acts only on the phase factors ${\cal V}$ and ${\cal V}^*$
on the right.

Introducing the Nambu-Gor'kov spinors
\begin{equation}
X_r \equiv \left( \begin{array}{c} 
                    \chi_r \\
                    \chi_{C r} 
                   \end{array}
            \right) \,\,\, , \,\,\,\,
\bar{X}_r \equiv ( \bar{\chi}_r \, , \, \bar{\chi}_{C r} )
\end{equation}
and the $2 N_c N_f \times 2 N_c N_f$-dimensional Maurer-Cartan one-form
\begin{equation} \label{Maurer2}
\Omega^\mu(x,y) \equiv -i \, \left( \begin{array}{cc}
                         \omega^\mu(x) & 0 \\
                         0 & \omega_C^\mu(x) 
                         \end{array}
                  \right)\, \delta^{(4)}(x-y) \,\, ,
\end{equation}
the quark (replica) partition function becomes
\begin{equation}
{\cal Z}_M[\Omega,\Phi^+,\Phi^-] \equiv
\int \prod_{r=1}^{M/2} {\cal D} \bar{X}_r \, {\cal D} X_r \; \exp 
\left\{ \sum_{r=1}^{M/2} 
\int_{x,y} \bar{X}_r(x) \,\left [\,  {\cal S}^{-1} (x,y)
+  \gamma_\mu \Omega^\mu(x,y) \, \right] \, X_r(y) 
\right\} \,\, , \label{Zquarks4}
\end{equation}
where
\begin{equation}
{\cal S}^{-1} \equiv 
\left( \begin{array}{cc}
            [G^+_0]^{-1} & \Phi^- \\
             \Phi^+ & [G^-_0]^{-1}
       \end{array} \right)\,\, .
\end{equation}

We are interested in the properties of the gluons, and thus
may integrate out the fermion fields. This
integration can be performed analytically, with the result
\begin{equation}
{\cal Z}_M[\Omega,\Phi^+,\Phi^-] \equiv
\left[ \,{\rm det} \left( {\cal S}^{-1} + \gamma_\mu \Omega^\mu 
\right) \, \right]^{M/2} \,\, . \label{Zquarks5}
\end{equation}
The determinant is to be taken over Nambu-Gor'kov,
color, flavor, spin, and space-time indices.
Finally, letting $M \rightarrow 1$, we obtain the QCD partition function 
(in the presence of meson, $\varphi_a$, and diquark, $\Phi^\pm$, source
fields)
\begin{equation} \label{ZQCD2}
{\cal Z}[\varphi,\Phi^+, \Phi^-]  =  
\int {\cal D} A \; \exp\left[ S_A 
+ \frac{1}{2} \, 
{\rm Tr} \ln \left({\cal S}^{-1} + \gamma_\mu  \Omega^\mu \right) \,
\right]\,\, .  
\end{equation}
Remembering that $\Omega^\mu$
is linear in $A^\mu_a$, cf.\ Eq.\ (\ref{Maurer2}) with (\ref{Maurer1}), 
in order to derive the gluon propagator it is
sufficient to expand the logarithm to second order in $\Omega^\mu$,
\begin{eqnarray} 
\frac{1}{2}\, 
{\rm Tr} \ln \left({\cal S}^{-1} + \gamma_\mu \Omega^\mu \right) \,
& \simeq & \frac{1}{2}\,{\rm Tr} \ln {\cal S}^{-1}
+ \frac{1}{2}\,{\rm Tr} \left( {\cal S}\, \gamma_\mu  \Omega^\mu \right)
- \frac{1}{4} {\rm Tr} \left(  
{\cal S} \, \gamma_\mu  \Omega^\mu \, {\cal S} \, \gamma_\nu  
\Omega^\nu \right) \nonumber \\
& \equiv & S_0[\Phi^+,\Phi^-] + S_1[\Omega,\Phi^+,\Phi^-] + 
S_2[\Omega,\Phi^+,\Phi^-]\,\,, \label{expandlog}
\end{eqnarray}
with obvious definitions for the $S_i$.
The quasiparticle propagator is
\begin{equation}
{\cal S} \equiv 
\left( \begin{array}{cc}
            G^+ & \Xi^- \\
             \Xi^+ & G^-
       \end{array} \right)\,\,,
\end{equation}
with
\begin{equation}
G^\pm = \left\{ [G_0^\pm]^{-1} - \Sigma^\pm \right\}^{-1}
\,\,\,\, , \,\,\,\,\,
\Sigma^\pm = \Phi^\mp \, G_0^\mp \, \Phi^\pm
\,\,\,\, ,\,\,\,\,\,
\Xi^\pm = - G_0^\mp \, \Phi^\pm \, G^\pm\,\, .
\end{equation}

To make further progress, we now expand
$\omega^\mu$ and $\omega_C^\mu $ to linear order in the meson fields,
\begin{mathletters} \label{linearomega}
\begin{eqnarray}
\omega^\mu & \simeq & g \, A^\mu_a \, T_a - \sum_{a=4}^7 
\left( \partial^\mu  \varphi_a \right)\, T_a - 
\frac{1}{\sqrt{3}}\, \left(\partial^\mu  \varphi_8\right)\, B\,\, , \\
\omega_C^\mu & \simeq & - g \, A^\mu_a \, T_a^T + \sum_{a=4}^7 
\left( \partial^\mu \varphi_a \right) \, T_a^T + 
\frac{1}{\sqrt{3}}\, \left( \partial^\mu  \varphi_8\right) \, B^T\,\, .
\end{eqnarray}
\end{mathletters}
The term $S_1$ in Eq.\ (\ref{expandlog}) is simply a
tadpole source term for the gluon fields. 
This term does not affect the gluon propagator, and thus
can be ignored in the following.

The quadratic term $S_2$ represents the contribution
of a fermion loop to the gluon self-energy.
Its computation proceeds by first taking the trace over Nambu-Gor'kov space,
\begin{eqnarray}
S_2 & = & -\frac{1}{4} \int_{x,y} {\rm Tr}_{c,f,s} \left[
G^+(x,y) \, \gamma_\mu \omega^\mu(y)\, G^+(y,x) \, \gamma_\nu \omega^\nu(x)
+ G^-(x,y) \, \gamma_\mu \omega_C^\mu(y)\, G^-(y,x) \, \gamma_\nu 
\omega_C^\nu(x) \right. \nonumber \\
&  & \hspace*{2.1cm} + \left.  
\Xi^+(x,y) \, \gamma_\mu \omega^\mu(y)\, \Xi^-(y,x) \, \gamma_\nu 
\omega_C^\nu(x)
+ \Xi^-(x,y) \, \gamma_\mu \omega_C^\mu(y)\, \Xi^+(y,x) \, \gamma_\nu
\omega^\nu(x) \right] \,\,. \label{S2}
\end{eqnarray}
The remaining trace runs only over color, flavor, and spin indices.
Using translational invariance, the propagators and fields are
now Fourier-transformed as
\begin{mathletters}
\begin{eqnarray}
G^\pm (x,y) & = & \frac{T}{V} \sum_K e^{-i K \cdot (x-y)} \, G^\pm(K)\,\, ,\\
\Xi^\pm (x,y) & = & \frac{T}{V} \sum_K e^{-i K \cdot (x-y)} \, 
\Xi^\pm(K)\,\, ,\\
\omega^\mu (x) & = &  \sum_P e^{-i P \cdot x} \, 
\omega^\mu(P)\,\, ,\\
\omega_C^\mu (x) & = &  \sum_P e^{-i P \cdot x} \, 
\omega_C^\mu(P)\,\, .
\end{eqnarray}
\end{mathletters}
Inserting this into Eq.\ (\ref{S2}), we arrive at Eq.\ (C16) 
of Ref.\ \cite{msw}, which in our notation reads
\begin{eqnarray}
S_2 & = & -\frac{1}{4} \sum_{K,P} {\rm Tr}_{c,f,s} \left[
G^+(K) \, \gamma_\mu \omega^\mu(P)\, G^+(K-P) \, \gamma_\nu \omega^\nu(-P)
+ G^-(K) \, \gamma_\mu \omega_C^\mu(P)\, G^-(K-P) \, \gamma_\nu 
\omega_C^\nu(-P) \right. \nonumber \\
&  & \hspace*{1.1cm} + \left.  
\Xi^+(K) \, \gamma_\mu \omega^\mu(P)\, \Xi^-(K-P) \, \gamma_\nu 
\omega_C^\nu(-P)
+ \Xi^-(K) \, \gamma_\mu \omega_C^\mu(P)\, \Xi^+(K-P) \, \gamma_\nu
\omega^\nu(-P) \right] \,\,. 
\end{eqnarray}
The remainder of the calculation is straightforward, but somewhat
tedious. First, insert the (Fourier-transform of the)
linearized version (\ref{linearomega})
for the fields $\omega^\mu$ and $\omega_C^\mu$. 
This produces a plethora of terms which are second order
in the gluon and meson fields, with coefficients that are
traces over color, flavor, and spin. Next, perform the color and flavor
traces in these coefficients. It turns out that some of them are 
identically zero, preventing the occurrence of terms
which mix gluons of adjoint colors 1, 2, and 3 (the unbroken
$SU(2)_c$ subgroup) among themselves and 
with the other gluon and meson fields. 
Furthermore, there are no terms mixing
the meson fields $\varphi_a,\, a=4, \ldots 7,$ with $\varphi_8$.
There are mixed terms between gluons and mesons with adjoint
color indices $4, \ldots, 7$, and between the gluon field
$A_8^\mu$ and the meson field $\varphi_8$.

Some of the mixed terms (those which mix gluons and mesons
of adjoint colors 4 and 5, as well as 6 and 7)
can be eliminated via a unitary transformation analogous to
the one employed in Ref.\ \cite{dhr2f}, Eq.\ (80).
Introducing the tensors
\begin{mathletters}
\begin{eqnarray}
\Pi^{\mu \nu}_{11} (P) & \equiv & \Pi^{\mu \nu}_{22} (P) \equiv 
\Pi^{\mu \nu}_{33} (P)  =  \frac{g^2}{2}  \, \frac{T}{V} 
\sum_K {\rm Tr}_{s} \left[ \gamma^\mu \, G^+ (K) \, \gamma^\nu 
\, G^+(K-P) + \gamma^\mu \, G^- (K) \, \gamma^\nu 
\, G^-(K-P) \right. \nonumber \\
&   & \left. \hspace*{5.1cm}
+\, \gamma^\mu \, \Xi^- (K) \, \gamma^\nu
\, \Xi^+(K-P) + \gamma^\mu \, \Xi^+ (K) \, \gamma^\nu 
\, \Xi^-(K-P)  \right] \,\,, \label{Pi11}
\end{eqnarray}
cf.\  Eq.\ (78a) of Ref.\ \cite{dhr2f},
\begin{eqnarray}
\Pi^{\mu \nu}_{44} (P) & \equiv & \Pi^{\mu \nu}_{66} (P) 
 =  \frac{g^2}{2}  \, \frac{T}{V} 
\sum_K {\rm Tr}_{s} \left[ \gamma^\mu \, G_0^+ (K) \, \gamma^\nu 
\, G^+(K-P) + \gamma^\mu \, G^- (K) \, \gamma^\nu 
\, G_0^-(K-P) \right]\,\, , \label{Pi44diag} 
\end{eqnarray}
cf.\ Eq.\ (83a) of Ref.\ \cite{dhr2f},
\begin{eqnarray}
\Pi^{\mu \nu}_{55} (P) & \equiv & \Pi^{\mu \nu}_{77} (P) 
 =  \frac{g^2}{2}  \, \frac{T}{V} 
\sum_K {\rm Tr}_{s} \left[ \gamma^\mu \, G^+ (K) \, \gamma^\nu 
\, G_0^+(K-P) + \gamma^\mu \, G_0^- (K) \, \gamma^\nu 
\, G^-(K-P) \right]\,\, . \label{Pi55diag}
\end{eqnarray}
cf.\ Eq.\ (83b) of Ref.\ \cite{dhr2f}, as well as
\begin{eqnarray}
\Pi^{\mu \nu}_{88} (P) & = & \frac{2}{3} \, \Pi_0^{\mu \nu}(P) 
+ \frac{1}{3} \, \tilde{\Pi}^{\mu \nu} (P) \,\, , \label{Pi88}\\
\tilde{\Pi}^{\mu \nu} (P) & = & \frac{g^2}{2} \, \frac{T}{V} 
\sum_K {\rm Tr}_{s} \left[ \gamma^\mu \, G^+ (K) \, \gamma^\nu 
\, G^+(K-P) + \gamma^\mu \, G^- (K) \, \gamma^\nu 
\, G^-(K-P) \right. \nonumber \\
&   & \left. \hspace*{1.8cm}
-\, \gamma^\mu \, \Xi^- (K) \, \gamma^\nu
\, \Xi^+(K-P) - \gamma^\mu \, \Xi^+ (K) \, \gamma^\nu 
\, \Xi^-(K-P)  \right] \,\,, \label{Pitilde}
\end{eqnarray}
cf.\ Eq.\ (78c) of Ref.\ \cite{dhr2f},
where $\Pi_0^{\mu \nu}$ is the gluon self-energy in a dense, but 
normal-conducting system,
\begin{equation}
\Pi_0^{\mu \nu} (P) =  \frac{g^2}{2}\, \frac{T}{V} \sum_K {\rm Tr}_{s} 
\left[\gamma^\mu \, G_0^+ (K)\, \gamma^\nu \,
G_0^+(K-P) + \gamma^\mu \,G_0^-(K)\,\gamma^\nu \,G_0^-(K-P) \right]\,\, ,
\label{Pi0}
\end{equation}
\end{mathletters}
cf.\ Eq.\ (27b) of Ref.\ \cite{dhr2f}, the final result can be written 
in the compact form (cf.\ Eq.\ (C19) of Ref.\ \cite{msw})
\begin{equation} \label{S2final}
S_2  =  - \frac{1}{2} \, \frac{V}{T} \, \sum_P \sum_{a=1}^8 
\left[A_\mu^a(-P) - \frac{i}{g}\, P_\mu\, \varphi^a(-P)\right]
\, \Pi^{\mu \nu}_{aa}(P) \, 
\left[A_\nu^a(P) + \frac{i}{g}\, P_\nu\, \varphi^a(P)\right] \,\, .
\end{equation}
In deriving Eq.\ (\ref{S2final}),
we have made use of
the transversality of the polarization tensor
in the normal-conducting phase, 
$\Pi_0^{\mu \nu}(P) \, P_\nu = P_\mu \, \Pi_0^{\mu \nu}(P)=0$.
Note that the tensors $\Pi^{\mu \nu}_{aa}$ for $a= 1, \, 2,$ 
and 3 are also transverse, but those for $a=4,\ldots,8$ are not.
This can be seen explicitly from the expressions given in Ref.\ 
\cite{dhrselfenergy}. 
The compact notation of Eq.\ (\ref{S2final}) is made possible
by the fact that $\varphi^a \equiv 0$ for $a = 1,2,3$, and because we
introduced the extra factor $1/\sqrt{3}$ in Eq.\ (\ref{phase})
as compared to Ref.\ \cite{msw}.

To make further progress, it is advantageous to 
tensor-decompose $\Pi^{\mu \nu}_{aa}$.
Various ways to do this are possible \cite{msw}; here we follow 
the notation of Ref.\ \cite{LeBellac}. 
First, define a projector onto the subspace parallel to $P^\mu$,
\begin{equation} \label{E}
{\rm E}^{\mu \nu} = \frac{P^\mu \, P^\nu}{P^2}\,\, .
\end{equation}
Then choose a vector orthogonal to $P^\mu$, for instance
\begin{equation}
N^\mu \equiv \left( \frac{p_0\, p^2}{P^2}, \frac{p_0^2\, {\bf p}}{P^2}
\right) \equiv \left(g^{\mu \nu} - {\rm E}^{\mu \nu}\right)\, f_\nu\,\, ,
\end{equation}
with $f^\mu = (0,{\bf p})$. Note that $N^2 = -p_0^2\,p^2/P^2$.
Now define the projectors
\begin{equation} \label{BCA}
{\rm B}^{\mu \nu} = \frac{N^\mu\, N^\nu}{N^2}\,\,\,\, , \,\,\,\,\,
{\rm C}^{\mu \nu} = N^\mu \, P^\nu + P^\mu\, N^\nu \,\,\,\, , \,\,\,\,\,
{\rm A}^{\mu \nu} = g^{\mu \nu} - {\rm B}^{\mu \nu} - {\rm E}^{\mu \nu} \,\, .
\end{equation}
Using the explicit form of $N^\mu$, one convinces oneself
that the tensor ${\rm A}^{\mu \nu}$ projects onto the spatially transverse
subspace orthogonal to $P^\mu$, 
\begin{equation}
{\rm A}^{00} = {\rm A}^{0i}=0\,\,\,\, , \,\,\,\,\,
{\rm A}^{ij} = - \left(\delta^{ij} - \hat{p}^i \, \hat{p}^j \right)\,\, .
\end{equation}
(Reference \cite{LeBellac} also uses the notation 
$P_T^{\mu \nu}$ for ${\rm A}^{\mu \nu}$.)
Consequently, the tensor ${\rm B}^{\mu \nu}$ projects onto the spatially
longitudinal subspace orthogonal to $P^\mu$,
\begin{equation}
{\rm B}^{00} = - \frac{p^2}{P^2} \,\,\,\, , \,\,\,\,\,
{\rm B}^{0i} = - \frac{p_0\, p^i}{P^2}\,\,\,\, ,\,\,\,\,
{\rm B}^{ij} = - \frac{p_0^2}{P^2}\,\hat{p}^i\,\hat{p}^j\,\, .
\end{equation}
(Reference \cite{LeBellac} also employs the notation
$P_L^{\mu \nu}$ for ${\rm B}^{\mu \nu}$.)
With these tensors, the gluon self-energy can be written in the form 
\begin{equation} \label{tensor}
\Pi^{\mu \nu}_{aa}(P) = \Pi^{\rm a}_{aa}(P) \, {\rm A}^{\mu \nu}
+ \Pi^{\rm b}_{aa}(P) \, {\rm B}^{\mu \nu} + \Pi^{\rm c}_{aa}(P)\, 
{\rm C}^{\mu \nu} + \Pi^{\rm e}_{aa}(P)\, {\rm E}^{\mu \nu}\,\, .
\end{equation}
The polarization functions $\Pi^{\rm a}_{aa},\, \Pi^{\rm b}_{aa}, \, 
\Pi^{\rm c}_{aa},$ and $\Pi^{\rm e}_{aa}$ 
can be computed by projecting the tensor
$\Pi^{\mu \nu}_{aa}$
onto the respective subspaces of the projectors (\ref{E}) and
(\ref{BCA}). Introducing the abbreviations
\begin{equation}
\Pi^t_{aa}(P) \equiv \frac{1}{2} \, 
\left( \delta^{ij} - \hat{p}^i\, \hat{p}^j \right) \, \Pi^{ij}_{aa}(P) 
\,\,\,\, ,\,\,\,\,\, 
\Pi^\ell_{aa}(P) \equiv \hat{p}_i \, \Pi^{ij}_{aa}(P)\, \hat{p}_j \,\, .
\end{equation}
these functions read
\begin{mathletters} \label{Pifunctions}
\begin{eqnarray}
\Pi^{\rm a}_{aa}(P) & = & \frac{1}{2}\, \Pi^{\mu \nu}_{aa}(P)\, 
{\rm A}_{\mu \nu} =  -  \Pi^t_{aa}(P) \,\, ,  \label{Pia} \\
\Pi^{\rm b}_{aa}(P) & = & \Pi^{\mu \nu}_{aa}(P)\, {\rm B}_{\mu \nu}
=  - \frac{p^2}{P^2} \, \left[ \Pi^{00}_{aa}(P)
+ 2\, \frac{p_0}{p}\, \Pi^{0i}_{aa}(P)\,\hat{p}_i
+ \frac{p_0^2}{p^2}\, \Pi^\ell_{aa}(P) \right] \,\, ,  \\
\Pi^{\rm c}_{aa}(P) & = & \frac{1}{2\, N^2 \, P^2}\, \Pi^{\mu \nu}_{aa}(P)\, 
{\rm C}_{\mu \nu}
= -\frac{1}{P^2}\, \left[ \Pi^{00}_{aa}(P)
+ \frac{p_0^2+p^2}{p_0\,p}\, \Pi^{0i}_{aa}(P)\,\hat{p}_i
+ \Pi^\ell_{aa}(P) \right] \,\, , \\
\Pi^{\rm e}_{aa}(P) & = & \Pi^{\mu \nu}_{aa}(P)\, {\rm E}_{\mu \nu}
= \frac{1}{P^2}\, \left[ p_0^2 \, \Pi^{00}_{aa}(P)
+ 2\,p_0\,p \, \Pi^{0i}_{aa}(P) \, \hat{p}_i
+ p^2 \, \Pi^\ell_{aa}(P) \right] \,\, .
\end{eqnarray}
\end{mathletters}
For the explicitly transverse tensor $\Pi^{\mu \nu}_{11}$,
the functions $\Pi^{\rm c}_{11} = \Pi^{\rm e}_{11} \equiv 0$.
The same holds for the HDL polarization tensor $\Pi_0^{\mu \nu}$.
For the other gluon colors $a=4, \ldots, 8$, the functions
$\Pi^{\rm c}_{aa}$ and $\Pi^{\rm e}_{aa}$ do not vanish.
Note that the dimensions of $\Pi^{\rm a}_{aa},\, \Pi^{\rm b}_{aa},$
and $\Pi^{\rm e}_{aa}$ are $[{\rm MeV}^2]$, while
$\Pi^{\rm c}_{aa}$ is dimensionless.

Now let us define the functions
\begin{equation} \label{functions}
A_{\perp\, \mu}^a(P) = {{\rm A}_\mu}^\nu\, A^a_\nu(P) 
\,\,\,\, , \,\,\,\,\,
A_\parallel^a(P) = \frac{ P^\mu \, A^a_\mu(P)}{P^2} 
\,\,\,\, , \,\,\,\,\,
A_N^a(P) = \frac{ N^\mu \, A^a_\mu(P)}{N^2}\,\, .
\end{equation}
Note that $A_\parallel^a(-P) = - P^\mu \, A^a_\mu(-P)/P^2$, and
$A_N^a(-P) = - N^\mu \, A^a_\mu(-P)/N^2$, since $N^\mu$ is odd
under $P \rightarrow -P$. The fields $A_\parallel^a(P)$
and $A_N^a(P)$ are dimensionless.
With the tensor decomposition (\ref{tensor}) and the 
functions (\ref{functions}), 
Eq.\ (\ref{S2final}) becomes
\begin{eqnarray}
S_2 & = &  -\frac{1}{2}\, \frac{V}{T} \sum_P \sum_{a=1}^8
\left\{ \frac{}{} A_{\perp\, \mu}^a(-P)\, \Pi^{\rm a}_{aa}(P)\, 
{\rm A}^{\mu \nu}\, A_{\perp\,\nu}^a(P)  
- A_N^a(-P) \, \Pi^{\rm b}_{aa}(P)\, N^2 \, A_N^a(P) 
\right.
\nonumber \\
&   &  - \left[A_\parallel^a(-P) + \frac{i}{g}\, \varphi^a(-P)\right]\,
\Pi^{\rm c}_{aa}(P) \,N^2 P^2 \, A_N^a(P) 
- A_N^a(-P) \, \Pi^{\rm c}_{aa}(P) \,N^2 P^2 \
\left[A_\parallel^a(P) + \frac{i}{g}\, \varphi^a(P)\right]
\nonumber  \\
&    & - \left. \left[A_\parallel^a(-P) + \frac{i}{g}\, \varphi^a(-P)\right]\,
\Pi^{\rm e}_{aa}(P) \, P^2 \, 
\left[A_\parallel^a(P) + \frac{i}{g}\, \varphi^a(P)\right] \right\}\,\,.
\label{decompose}
\end{eqnarray}

In any spontaneously broken gauge theory, 
the excitations of the condensate mix with 
the gauge fields corresponding to the broken generators of the 
underlying gauge group. The mixing occurs in the
components orthogonal to the spatially transverse degrees of freedom,
{\it i.e.}, for the spatially longitudinal fields, $A_N^a$, and 
the fields parallel to $P^\mu$, $A_\parallel^a$.
For the two-flavor color superconductor, these components
mix with the meson fields for gluon colors $4, \ldots, 8$. 
The mixing is particularly evident in Eq.\ (\ref{decompose}).

The terms mixing mesons and gauge fields can be
eliminated by a suitable choice of gauge.
The gauge to accomplish this goal is the 't~Hooft gauge.
The ``unmixing'' procedure of mesons and gauge fields
consists of two steps.
First, we eliminate the terms in Eq.\ (\ref{decompose}) which
mix $A_N^a$ and $A_\parallel^a$.
This is achieved by substituting
\begin{equation} \label{sub}
\hat{A}_\parallel^a(P) = A_\parallel^a(P) +
\frac{\Pi^{\rm c}_{aa}(P)\, N^2}{\Pi^{\rm e}_{aa}(P)} \, A_N^a(P)\,\, .
\end{equation}
(We do not perform this substitution for $a=1,2,3$; for these
gluon colors $\Pi^{\rm c}_{aa}\equiv 0$, such that
there are no terms in Eq.\ (\ref{decompose}) which mix
$A_\parallel^a$ and $A_N^a$).
This shift of the gauge field component $A_\parallel^a$ is
completely innocuous for the following reasons. First, the Jacobian 
$\partial(\hat{A}_\parallel, A_N)/\partial(A_\parallel, A_N)$ 
is unity, so the measure of the functional integral over gauge 
fields is not affected. Second, the only other term in the gauge 
field action, which is quadratic in
the gauge fields and thus relevant for the derivation
of the gluon propagator, is the free field action
\begin{equation} \label{S0}
S_{F^2}^{(0)} 
\equiv - \frac{1}{2} \, \frac{V}{T} \sum_P \sum_{a=1}^8 A_\mu^a(-P) \,
\left(P^2\, g^{\mu \nu} - P^{\mu}\, P^{\nu} \right) \, A_\nu^a(P)
\equiv - \frac{1}{2} \, \frac{V}{T} \sum_P \sum_{a=1}^8 A_\mu^a(-P) \, P^2 \,
\left({\rm A}^{\mu \nu} + {\rm B}^{\mu \nu} \right) \, A_\nu^a(P)\,\, ,
\end{equation}
and it does not contain the parallel components $A_\parallel^a(P)$.
It is therefore also not affected by the shift of variables
(\ref{sub}).

After renaming $\hat{A}_\parallel^a \rightarrow A_\parallel^a$, 
the final result for $S_2$ reads:
\begin{eqnarray}
S_2 & = & -\frac{1}{2}\, \frac{V}{T} \sum_P \sum_{a=1}^8
\left\{ \frac{}{} A_{\perp\, \mu}^a(-P)\, \Pi^{\rm a}_{aa}(P)\, A^{\mu \nu}\,
A_{\perp\,\nu}^a(P) 
 -  A_N^a(-P) \,\hat{\Pi}^{\rm b}_{aa}(P) \, N^2 \, A_N^a(P) \right.
\nonumber \\
&   & \hspace*{1.85cm}
- \left. \left[A_\parallel^a(-P) + \frac{i}{g}\, \varphi^a(-P)\right]\,
\Pi^{\rm e}_{aa}(P) \, P^2 \, 
\left[A_\parallel^a(P) + \frac{i}{g}\, \varphi^a(P)\right] \right\}\,\,,
\label{S2finalunmix}
\end{eqnarray}
where we introduced
\begin{equation} \label{hatPib}
\hat{\Pi}^{\rm b}_{aa}(P) \equiv  \Pi^{\rm b}_{aa}(P)
- \frac{\left[\Pi^{\rm c}_{aa}(P)\right]^2 N^2 P^2}{\Pi^{\rm e}_{aa}(P)} \,\, .
\end{equation}
The 't~Hooft gauge fixing term is now chosen to eliminate the
mixing between $A_\parallel^a$ and $\varphi^a$:
\begin{equation}
S_{\rm gf} =  \frac{1}{2 \, \lambda} \,\frac{V}{T} 
\sum_P \sum_{a=1}^8
\left[ P^2\, A_\parallel^a(-P) -
\lambda\, \frac{i}{g} \, \Pi^{\rm e}_{aa}(P)\, \varphi^a(-P) \right] \,
\left[ P^2\, A_\parallel^a(P) - 
\lambda\, \frac{i}{g}\, \Pi^{\rm e}_{aa}(P)\, \varphi^a(P) \right] \,\, .
\label{L_gf}
\end{equation}
This gauge condition is non-local in coordinate space, 
which seems peculiar, but poses no problem in momentum space.
Note that $P^2\, A_\parallel^a(P) \equiv P^\mu \, A_\mu^a(P)$. Therefore,
in various limits the choice of gauge (\ref{L_gf}) corresponds 
to covariant gauge,
\begin{equation}
S_{\rm cg} = \frac{1}{2 \, \lambda} \,\frac{V}{T} 
\sum_P \sum_{a=1}^8  A_\mu^a(-P)\, P^\mu\, P^\nu \, A_\nu^a(P) \,\, .
\label{L_cg}
\end{equation}
The first limit we consider is $T, \mu \rightarrow 0$, 
{\it i.e.\/} the vacuum. 
Then, $\Pi^{\rm e}_{aa} \equiv 0$, and Eq.\ (\ref{L_gf}) becomes (\ref{L_cg}).
The second case is the limit of large 4-momenta, $P \rightarrow \infty$.
As shown in Ref.\ \cite{dhrselfenergy}, in this region of phase space 
the effects from a color-superconducting condensate on the
gluon polarization tensor are negligible. In other words,
the gluon polarization tensor approaches the HDL limit.
The physical reason is that gluons with large momenta do not
see quark Cooper pairs as composite objects, but resolve the
individual color charges inside the pair. Consequently,
$\Pi^{\rm e}_{aa}(P) \, P^2 \rightarrow P_\mu \, \Pi^{\mu \nu}_0(P)\,
P_\nu \equiv 0$ for $P \rightarrow \infty$
and, for large $P$, the individual terms in the sum over $P$ in 
Eqs.\ (\ref{L_gf}) and (\ref{L_cg}) agree.
Finally, for gluon colors $a=1,2,3$,
$\Pi^{\rm e}_{aa} \equiv 0$, since the self-energy $\Pi^{\mu \nu}_{11}$
is transverse. Thus, for $a=1,2,3$ the terms 
in Eqs.\ (\ref{L_gf}) and (\ref{L_cg}) are identical.

The decoupling of mesons and gluon degrees of freedom
becomes obvious once we add (\ref{L_gf})
to (\ref{S2finalunmix}) and (\ref{S0}),
\begin{eqnarray}
S_{F^2}^{(0)} + S_2 + S_{\rm gf} 
& = & -\frac{1}{2} \, \frac{V}{T} \sum_P \sum_{a=1}^8
\left\{ \frac{}{} A_{\perp\, \mu}^a(-P)\, 
\left[ P^2 + \Pi^{\rm a}_{aa}(P) \right]\, 
{\rm A}^{\mu \nu}\, A_{\perp\,\nu}^a(P)  \right. \nonumber \\
&   & \hspace*{2.1cm} - \; A_N^a(-P) \, \left[ P^2 + \hat{\Pi}^{\rm b}_{aa}(P)
\right] \, N^2 \, A_N^a(P) \nonumber \\
&   &  \hspace*{2.1cm} - \;  
A_\parallel^a(-P) \, \left[ \frac{1}{\lambda}\, P^2
+ \Pi^{\rm e}_{aa}(P) \right]\, P^2 \, A_\parallel^a(P) \nonumber \\
&    & \hspace*{2.1cm} + \left. \frac{\lambda}{g^2}\, \varphi^a(-P)\,
\left[ \frac{1}{\lambda}\, P^2 + \Pi^{\rm e}_{aa}(P) \right] \, 
\Pi^{\rm e}_{aa}(P)\, \varphi^a(P) \right\} \,\, .
\label{SgfS2}
\end{eqnarray}
Consequently, the inverse gluon propagator is
\begin{equation}
{\Delta^{-1}}^{\mu \nu}_{aa}(P)  =  
\left[ P^2 + \Pi^{\rm a}_{aa}(P) \right]\, {\rm A}^{\mu \nu} 
+ \left[ P^2 + \hat{\Pi}^{\rm b}_{aa}(P) \right] \, {\rm B}^{\mu \nu} + 
\left[ \frac{1}{\lambda}\, P^2 + \Pi^{\rm e}_{aa}(P) \right] \, 
{\rm E}^{\mu \nu}\,\, .
\end{equation}
Inverting this as discussed in Ref.\ \cite{LeBellac}, one obtains
the gluon propagator for gluons of color $a$,
\begin{equation} \label{glueprop}
\Delta^{\mu \nu}_{aa}(P) = \frac{1}{P^2 + \Pi^{\rm a}_{aa}(P)}\,
{\rm A}^{\mu \nu} + 
\frac{1}{P^2 + \hat{\Pi}^{\rm b}_{aa}(P)}\, {\rm B}^{\mu \nu}
+ \frac{\lambda}{P^2 + \lambda\, \Pi^{\rm e}_{aa}(P)} \, 
{\rm E}^{\mu \nu}\,\, .
\end{equation}
For any $\lambda \neq 0$, the gluon propagator contains 
unphysical contributions parallel to $P^\mu$, which have to be cancelled
by the corresponding Faddeev-Popov ghosts when computing
physical observables. Only for $\lambda = 0$ these contributions
vanish and the gluon propagator is explicitly transverse,
{\it i.e.}, $P_\mu\, \Delta^{\mu \nu}_{aa}(P) = 
\Delta^{\mu \nu}_{aa}(P)\,P_\nu = 0$. Also, in this case
the ghost propagator is independent of the chemical potential $\mu$.
The contribution of Fadeev-Popov ghosts to the gluon polarization
tensor is then $\sim g^2\,T^2$ and thus negligible at $T=0$.
We shall therefore focus on this particular choice for
the gauge parameter in the following.
Note that for $\lambda = 0$, the inverse meson field
propagator is 
\begin{equation} \label{NGbosons}
D^{-1}_{aa}(P) \equiv \Pi^{\rm e}_{aa}(P)\, P^2
= P_\mu \, \Pi^{\mu \nu}_{aa}(P)\, P_\nu \,\, ,
\end{equation} 
and the dispersion relation for the mesons follows from the condition
$D^{-1}_{aa}(P)=0$, as demonstrated in Ref.\ \cite{zarembo} for
a three-flavor color superconductor in the color-flavor-locked
phase.

The gluon propagator for transverse and longitudinal
modes can now be read off Eq.\ (\ref{glueprop}) as coefficients
of the corresponding tensors ${\rm A}^{\mu \nu}$ (the projector
onto the spatially transverse subspace orthogonal to $P^\mu$)
and ${\rm B}^{\mu \nu}$ (the projector onto the spatially
longitudinal subspace orthogonal to $P^\mu$).
For the transverse modes one has \cite{LeBellac}
\begin{equation} \label{transverse}
\Delta^t_{aa}(P) \equiv \frac{1}{P^2 + \Pi^{\rm a}_{aa}(P)}
= \frac{1}{P^2 - \Pi^t_{aa}(P)}\,\, ,
\end{equation}
where we used Eq.\ (\ref{Pia}). Multiplying the
coefficient of ${\rm B}^{\mu \nu}$ in Eq.\ (\ref{glueprop})
with the standard factor $-P^2/p^2$ \cite{LeBellac}, one
obtains for the longitudinal modes
\begin{equation} \label{longitudinal}
\hat{\Delta}^{00}_{aa}(P) \equiv - \frac{P^2}{p^2} 
\, \frac{1}{P^2 + \hat{\Pi}^{\rm b}_{aa}(P)}
= - \frac{1}{p^2 - \hat{\Pi}^{00}_{aa}(P)}\,\, ,
\end{equation}
where the longitudinal gluon self-energy
\begin{equation} \label{hatPi00aa}
\hat{\Pi}^{00}_{aa}(P) \equiv  p^2 \,
\frac{\Pi^{00}_{aa}(P)\,\Pi^\ell_{aa}(P) - 
\left[ \Pi^{0i}_{aa}(P) \, \hat{p}_i \right]^2 }{
p_0^2 \, \Pi^{00}_{aa}(P) + 2\, p_0\,p\, \Pi^{0i}_{aa}(P) \, \hat{p}_i
+ p^2 \, \Pi^\ell_{aa}(P) } 
\end{equation}
follows from the definition of $\hat{\Pi}^{\rm b}_{aa}$, Eq.\
(\ref{hatPib}), and the relations (\ref{Pifunctions}).
The longitudinal gluon propagator
$\hat{\Delta}^{00}_{aa}$ {\em must not be confused\/} with the
the $00$-component of $\Delta^{\mu \nu}_{aa}$. 
We deliberately use this (slightly ambiguous)
notation to facilitate the comparison
of our new and correct results with those of 
Ref.\ \cite{dhrselfenergy}, which were partially incorrect.
The results of that paper were derived in Coulomb gauge, where 
the $00$-component of the propagator is indeed {\em identical\/} to
the longitudinal propagator (\ref{longitudinal}).
We were not able to find a 't~Hooft gauge that converged
to the Coulomb gauge in the various limits discussed above,
and consequently had to base our discussion on the
covariant gauge (\ref{L_cg}) as limiting case of Eq.\ (\ref{L_gf}).

To summarize this section, we have computed the gluon 
propagator for gluons in a two-flavor color superconductor. 
Due to condensation of quark Cooper pairs, the $SU(3)_c$
gauge symmetry is spontaneously broken to $SU(2)_c$, 
leading to the appearance of five Nambu-Goldstone bosons.
In general, these bosons mix with some components of the
gauge fields corresponding to the broken generators.
To ``unmix'' them we have used a form of 't~Hooft gauge which
smoothly converges to covariant gauge in the vacuum, as well as
for large gluon momenta, and when the gluon polarization tensor 
is explicitly transverse.
Finally, choosing the gauge fixing parameter $\lambda=0$ we
derived the gluon propagator for transverse, Eq.\ (\ref{transverse}), 
and longitudinal modes, Eq.\ (\ref{longitudinal}) with
(\ref{hatPi00aa}).

\section{Spectral properties of the eighth gluon} \label{III}

In this section, we explicitly compute the spectral
properties of the eighth gluon.
We shall confirm the results of Ref.\ \cite{dhrselfenergy} 
for the transverse mode and amend those for
the longitudinal mode, which have not been correctly computed
in Ref.\ \cite{dhrselfenergy}. In particular, we shall show
that the plasmon dispersion relation now has
the correct behavior $p_0 \rightarrow m_g$ as $p \rightarrow 0$.
Furthermore, the longitudinal spectral density vanishes for 
gluon energies and momenta
located on the dispersion branch of the Nambu-Goldstone bosons,
{\it i.e.}, for energies and momenta given by the roots
of Eq.\ (\ref{NGbosons}). For the eighth gluon,
this condition can be written in the form
$P_\mu\, \tilde{\Pi}^{\mu \nu}(P)\, P_\nu = 0$ \cite{gusyshov,zarembo}, 
since the HDL self-energy is transverse, 
$P_\mu\, \Pi^{\mu \nu}_0(P)\, P_\nu \equiv 0$.

\subsection{Polarization tensor}

We first compute the polarization tensor for the transverse
and longitudinal components of the eighth gluon.
To this end, it is convenient to rewrite the
longitudinal gluon self-energy (\ref{hatPi00aa})
in the form
\begin{eqnarray} \label{Pi0088}
\hat{\Pi}^{00}_{88}(P) & \equiv &
\frac{2}{3} \, \Pi_0^{00}(P) + \frac{1}{3}\, \hat{\Pi}^{00}(P) \,\, , \\
\hat{\Pi}^{00}(P) & \equiv & p^2 \,
\frac{\tilde{\Pi}^{00}(P)\,\tilde{\Pi}^\ell(P) - 
\left[ \tilde{\Pi}^{0i}(P) \, \hat{p}_i \right]^2 }{
p_0^2 \, \tilde{\Pi}^{00}(P) + 2\, p_0\,p\, \tilde{\Pi}^{0i}(P) \, \hat{p}_i
+ p^2 \, \tilde{\Pi}^\ell(P) } \,\, , \label{hatPi00}
\end{eqnarray}
with $\tilde{\Pi}^\ell (P) \equiv \hat{p}_i \, \tilde{\Pi}^{ij}(P)\,
\hat{p}_j$.

Let us now explicitly compute the polarization functions.
As in Ref.\ \cite{dhrselfenergy} we take $T=0$, and we
shall use the identity
\begin{equation} \label{ident}
\frac{1}{x+i \eta} \equiv {\cal P}\, \frac{1}{x} - i \pi \, \delta(x)\,\, ,
\end{equation}
where ${\cal P}$ stands for the principal value description,
in order to decompose the polarization tensor into real and imaginary
parts. The imaginary parts can then be straightforwardly computed,
while the real parts are computed from the dispersion integral
\begin{mathletters}
\begin{equation} \label{dispint}
{\rm Re} \, \Pi(p_0,{\bf p}) \equiv \frac{1}{\pi} \, {\cal P}
\int_{- \infty}^{\infty} d \omega\, \frac{{\rm Im}\, \Pi(\omega,{\bf p})}{
\omega - p_0} + C\,\, ,
\end{equation}
where $C$ is a (subtraction) constant.
If ${\rm Im}\, \Pi(\omega, {\bf p})$ is an odd function of $\omega$,
${\rm Im}\, \Pi(-\omega, {\bf p}) = - {\rm Im}\, \Pi(\omega, {\bf p})$,
Eq.\ (\ref{dispint}) becomes Eq.\ (39) of Ref.\ \cite{dhrselfenergy},
\begin{equation} \label{odd}
{\rm Re} \, \Pi(p_0,{\bf p}) \equiv \frac{1}{\pi} \, {\cal P}
\int_0^{\infty} d \omega\, {\rm Im}\, \Pi_{\rm odd}(\omega,{\bf p}) \,
\left(\frac{1}{\omega+p_0} + \frac{1}{\omega - p_0} \right) + C\,\, ,
\end{equation}
and if it is an even function of $\omega$,
${\rm Im}\, \Pi(-\omega, {\bf p}) = {\rm Im}\, \Pi(\omega, {\bf p})$,
we have instead
\begin{equation} \label{even}
{\rm Re} \, \Pi(p_0,{\bf p}) \equiv \frac{1}{\pi} \, {\cal P}
\int_0^{\infty} d \omega\, {\rm Im}\, \Pi_{\rm even}(\omega,{\bf p}) \,
\left(\frac{1}{\omega-p_0} - \frac{1}{\omega + p_0} \right) + C\,\, ,
\end{equation}
\end{mathletters}

Since the polarization tensor for the transverse gluon modes,
$\Pi^t_{88} \equiv \frac{2}{3}\, \Pi_0^t + \frac{1}{3}\, \tilde{\Pi}^t$, 
has already been computed in Ref.\ \cite{dhrselfenergy},
we just cite the results. The imaginary part of the
transverse HDL polarization function reads (cf.\ Eq.\ (22b)
of Ref.\ \cite{dhrselfenergy})
\begin{mathletters}
\begin{equation}
{\rm Im}\, \Pi^t_0(P) = - \pi\, \frac{3}{4}\, m_g^2 \, 
\frac{p_0}{p}\, \left(1- \frac{p_0^2}{p^2} \right) \, \theta(p-p_0)\,\, .
\end{equation}
The corresponding real part is computed from Eq.\ (\ref{odd}), with
the result (cf.\ Eqs.\ (40b) and (41) of Ref.\ \cite{dhrselfenergy})
\begin{equation}
{\rm Re}\, \Pi^t_0(P) = \frac{3}{2}\, m_g^2\, \left[
\frac{p_0^2}{p^2} + \left( 1- \frac{p_0^2}{p^2} \right) \,
\frac{p_0}{2\, p} \, \ln \left| \frac{p_0 + p}{p_0 - p} \right| \,
\right]\,\, .
\end{equation}
\end{mathletters}
We have used the fact that the value of the subtraction constant is
$C^t_0=m_g^2$, which can be derived from comparing
a direct calculation of ${\rm Re}\, \Pi^t_0$ using Eq.\ (19b)
of Ref.\ \cite{dhrselfenergy} with the above computation via the
dispersion formula (\ref{odd}).

The imaginary part of the tensor $\tilde{\Pi}^t$ is given by
(cf.\ Eq.\ (36) of Ref.\ \cite{dhrselfenergy})
\begin{eqnarray}
\lefteqn{{\rm Im}\, \tilde{\Pi}^t(P)  =
- \pi\, \frac{3}{4}\, m_g^2 \, \theta(p_0 - 2\, \phi)\, \frac{p_0}{p}
\left\{ \frac{}{}
\theta(E_p - p_0) \, \left[ \left( 1 - \frac{p_0^2}{p^2}\, (1+s^2)
\right) \, {\bf E}(t) - s^2 \,\left( 1- 2\, \frac{p_0^2}{p^2} \right) \, 
{\bf K}(t) \right]
\right.} \nonumber \\
& + & \left. \theta(p_0 - E_p) \left[ \left( 1 - \frac{p_0^2}{p^2}\, (1+s^2)
\right) \, E(\alpha,t) - \left( 1- \frac{p_0^2}{p^2} \right)\,
\frac{p}{p_0}\, \sqrt{1 - \frac{4\, \phi^2}{p_0^2 - p^2}} 
-  s^2 \,\left( 1- 2\, \frac{p_0^2}{p^2} \right)\, F(\alpha,t) \right]
\right\}\,\, ,
\end{eqnarray}
where $\phi$ is the value of the color-superconducting gap, 
$E_p = \sqrt{p^2 + 4 \phi^2}$,
$t = \sqrt{1-4\phi^2/p_0^2}$, $s^2 = 1 - t^2$, $\alpha = \arcsin [p/(t p_0)]$,
and $F(\alpha,t)$, $E(\alpha,t)$ are elliptic integrals of the
first and second kind, while ${\bf K}(t) \equiv F( \pi/2, t)$ and
${\bf E}(t) \equiv E( \pi/2,t)$ are the corresponding
complete elliptic integrals.
The real part is again computed from Eq.\ (\ref{odd}). The
integral has to be done numerically, see Appendix A of Ref.\ 
\cite{dhrselfenergy} for details. The subtraction constant is,
for reasons discussed at length in Ref.\ \cite{dhrselfenergy},
identical to the one in the HDL limit, $C^t \equiv C^t_0 = m_g^2$.
Finally, taking the linear combination 
$\Pi^t_{88} \equiv \frac{2}{3}\, \Pi_0^t + \frac{1}{3}\, \tilde{\Pi}^t$
completes the calculation of the transverse polarization function $\Pi^t_{88}$.

In order to compute the polarization function for the longitudinal
gluon, $\hat{\Pi}^{00}_{88}$, we have to know the functions
$\Pi_0^{00}(P)$, $\tilde{\Pi}^{00}(P)$, $\tilde{\Pi}^{0i}(P)\, \hat{p}_i$,
and $\tilde{\Pi}^\ell(P)$. The first two functions, 
$\Pi_0^{00}(P)$ and $\tilde{\Pi}^{00}(P)$
have also been computed in Ref.\ \cite{dhrselfenergy}. The imaginary part
of the longitudinal HDL polarization function is (cf.\ Eq.\ (22a)
of Ref.\ \cite{dhrselfenergy})
\begin{mathletters}
\begin{equation}
{\rm Im}\, \Pi_0^{00}(P) = - \pi\, \frac{3}{2}\, m_g^2 \, 
\frac{p_0}{p} \, \theta(p-p_0)\,\, .
\end{equation}
The real part is computed from Eq.\ (\ref{odd}), with
the result (cf.\ Eqs.\ (40a) and (41) of Ref.\ \cite{dhrselfenergy})
\begin{equation}
{\rm Re}\, \Pi^{00}_0(P) = - 3\, m_g^2\, \left( 1-
\frac{p_0}{2\, p} \, \ln \left| \frac{p_0 + p}{p_0 - p} \right| \,
\right)\,\, . \label{RePi000}
\end{equation}
\end{mathletters}
Here, the subtraction constant is $C^{00}_0 = 0$.

The imaginary part of the function $\tilde{\Pi}^{00}$ is
(cf.\ Eq.\ (35) of Ref.\ \cite{dhrselfenergy})
\begin{equation}
{\rm Im}\, \tilde{\Pi}^{00}(P)  =
- \pi\, \frac{3}{2}\, m_g^2 \, \theta(p_0 - 2\, \phi)\, \frac{p_0}{p}
\left\{ \frac{}{}
\theta(E_p - p_0) \, {\bf E}(t) + 
\theta(p_0 - E_p) \left[ E(\alpha,t) - 
\frac{p}{p_0}\, \sqrt{1 - \frac{4\, \phi^2}{p_0^2 - p^2}} 
\right] \right\}\,\, .
\end{equation}
The real part is computed from Eq.\ (\ref{odd}), with the
subtraction constant $C^{00} \equiv C^{00}_0 = 0$. Again, the
integral has to be done numerically.

It remains to compute the functions $\tilde{\Pi}^{0i}(P)\, \hat{p}_i$
and $\tilde{\Pi}^\ell(P)$. First, one performs the spin traces
in Eq.\ (\ref{Pitilde}) to obtain Eqs.\ (102b) and (102c)
of Ref.\ \cite{dhr2f}. Then, taking $T=0$, 
\begin{mathletters} \label{Pi0iPil}
\begin{eqnarray}
\tilde{\Pi}^{0i}(P)\, \hat{p}_i & = & 
\frac{g^2}{2} \int \frac{d^3 {\bf k}}{(2 \pi)^3} \,
\sum_{e_1,e_2= \pm} \left( e_1\, \hat{\bf k}_1 \cdot {\bf p}
+ e_2\, \hat{\bf k}_2 \cdot {\bf p} \right)\,
\left( \frac{\xi_2}{2 \epsilon_2} - \frac{\xi_1}{2 \epsilon_1} \right)
\nonumber \\
&     & \hspace*{2.6cm} \times \,
\left( \frac{1}{p_0 + \epsilon_1 + \epsilon_2 + i \eta}
+ \frac{1}{p_0 - \epsilon_1 - \epsilon_2+ i \eta} \right) \,\, , \\
\tilde{\Pi}^\ell(P) & = & 
- \frac{g^2}{2} \int \frac{d^3 {\bf k}}{(2 \pi)^3} \,
\sum_{e_1,e_2= \pm} \left[ \left(1- e_1e_2\, \hat{\bf k}_1 \cdot {\bf k}_2
\right) + 2\, e_1 e_2\, \hat{\bf k}_1 \cdot {\bf p}\;
\hat{\bf k}_2 \cdot {\bf p} \right]\,
\frac{\epsilon_1 \epsilon_2 - \xi_1 \xi_2 - \phi_1 \phi_2}{2 \, 
\epsilon_1 \epsilon_2}
\nonumber \\
&     & \hspace*{2.6cm} \times \,
\left( \frac{1}{p_0 + \epsilon_1 + \epsilon_2 + i \eta}
- \frac{1}{p_0 - \epsilon_1 - \epsilon_2+ i \eta} \right)  \,\, , 
\end{eqnarray}
\end{mathletters}
where ${\bf k}_{1,2} = {\bf k} \pm {\bf p}/2$,
$\phi_i \equiv \phi^{e_i}_{{\bf k}_i}$ is the
gap function for quasiparticles ($e_i = +1$) or
quasi-antiparticles ($e_i = -1$) with momentum ${\bf k}_i$,
$\xi_i \equiv e_ik_i - \mu$, and $\epsilon_i \equiv \sqrt{\xi_i^2 
+ \phi_i^2}$. 

One now repeats the steps discussed in detail in Section II.A of
Ref.\ \cite{dhrselfenergy} to obtain (for $p_0 \geq 0$)
\begin{mathletters} \label{Pi0iPil2}
\begin{eqnarray}
{\rm Im}\, \tilde{\Pi}^{0i}(P)\, \hat{p}_i & = & 
\pi\, \frac{3}{2}\, m_g^2 \, \theta(p_0 - 2\, \phi)\, \frac{p_0^2}{p^2}
\left\{ \frac{}{}
\theta(E_p - p_0) \, \left[ {\bf E}(t) - s^2 \, {\bf K}(t) \right]
\right. \nonumber \\
&    & \hspace*{3.2cm}
+ \left. \theta(p_0 - E_p) \left[ E(\alpha,t) - \frac{p}{p_0}\, \sqrt{1
- \frac{4\, \phi^2}{p_0^2 - p^2}} - s^2 \, F(\alpha,t) \right]
\right\}\,\, , \\
{\rm Im}\, \tilde{\Pi}^\ell(P) & = & 
- \pi\, \frac{3}{2}\, m_g^2 \, \theta(p_0 - 2\, \phi)\, \frac{p_0^3}{p^3}
\left\{ \frac{}{}
\theta(E_p - p_0) \, \left[ (1+s^2)\, {\bf E}(t) 
- 2\, s^2 \, {\bf K}(t) \right]
\right. \nonumber \\
&    & \hspace*{2.5cm}
+ \left. \theta(p_0 - E_p) \left[ (1+s^2)\, E(\alpha,t) - 
\frac{p}{p_0}\, \sqrt{1 - \frac{4\, \phi^2}{p_0^2 - p^2}} 
- 2\, s^2 \, F(\alpha,t) \right]
\right\}\,\, .
\end{eqnarray}
\end{mathletters}
One observes that in the limit $\phi \rightarrow 0$, the functions
(\ref{Pi0iPil2}) approach the HDL result
\begin{mathletters}
\begin{eqnarray}
{\rm Im}\, \Pi_0^{0i}(P)\, \hat{p}_i & = & 
\pi\, \frac{3}{2}\, m_g^2 \, \frac{p_0^2}{p^2} \, \theta(p-p_0)\,\, , \\
{\rm Im}\, \Pi_0^\ell(P) & = & 
- \pi\, \frac{3}{2}\, m_g^2 \, \frac{p_0^3}{p^3} \, \theta(p-p_0)\,\,.
\end{eqnarray}
\end{mathletters}
Applying Eq.\ (\ref{ident}) to Eqs.\ (\ref{Pi0iPil}) we immediately see
that the imaginary part of $\tilde{\Pi}^{0i}(P)\, \hat{p}_i$
is {\em even}, while that of $\tilde{\Pi}^\ell(P)$ is {\em odd}.
Thus, in order to compute the real part of $\tilde{\Pi}^{0i}(P)\, \hat{p}_i$,
we have to use Eq.\ (\ref{even}), while the real part of $\tilde{\Pi}^\ell(P)$
has to be computed from Eq.\ (\ref{odd}). When
implementing the numerical procedure discussed in Appendix A 
of Ref.\ \cite{dhrselfenergy} for the integral in Eq.\ (\ref{even}), 
one has to modify Eq.\ (A1) of Ref.\ \cite{dhrselfenergy} appropriately.

Finally, one has to determine the
values of the subtraction constants $C^{0i}$ and $C^\ell$.
We again use the fact that
$C^{0i} \equiv C^{0i}_0$ and $C^\ell \equiv C^\ell_0$,
where the index ``0'' refers to the HDL limit.
The corresponding constants 
are determined by first computing ${\rm Re}\, \Pi_0^{0i}(P)\,
\hat{p}_i$ and ${\rm Re}\, \Pi_0^\ell(P)$ from the
dispersion formulas (\ref{odd}) and (\ref{even}). The result of
this calculation is then compared to that of
a direct computation using, for instance,
the result (\ref{RePi000}) for ${\rm Re}\, \Pi_0^{00}(P)$
and then inferring ${\rm Re}\, \Pi_0^{0i}(P) \, \hat{p}_i$ and 
${\rm Re}\, \Pi_0^\ell(P)$ from the transversality of $\Pi_0^{\mu \nu}$.
The result is 
$C^{0i} \equiv C^{0i}_0 = 0$ and $C^{\ell}\equiv C^\ell_0 = m_g^2$.

\begin{figure}
\hspace*{2cm}
\mbox{\epsfig{figure=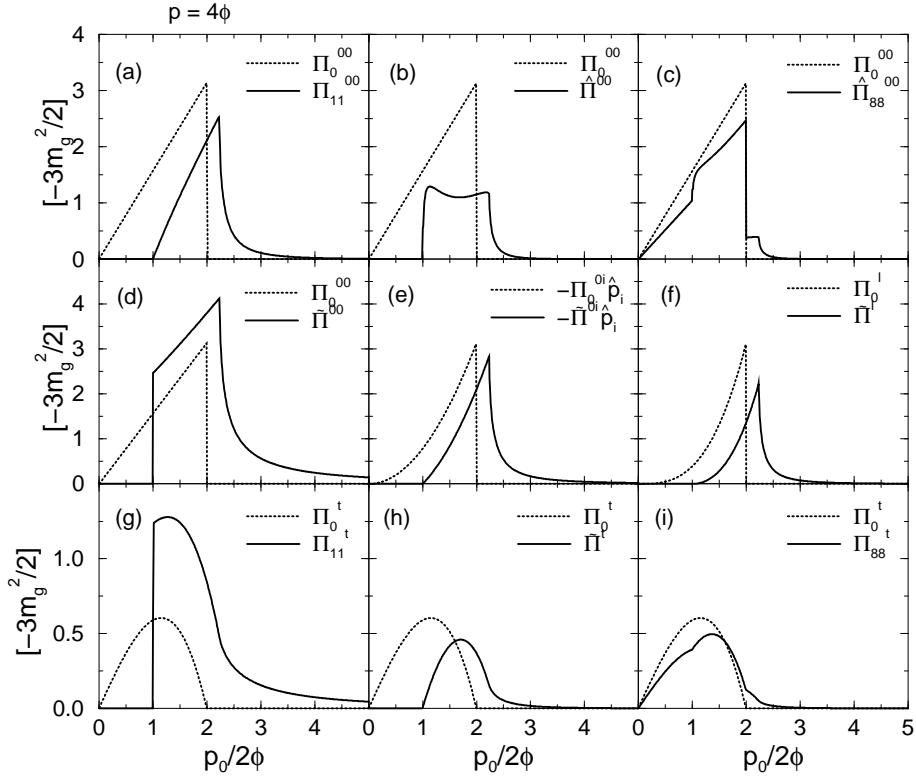,height=12cm,angle=270}}
\caption{Imaginary parts of the polarization tensors
in a two-flavor color superconductor (solid lines)
as a function of gluon energy $p_0$ for fixed gluon momentum $p= 4\, \phi$. 
(a) ${\rm Im}\, \Pi^{00}_{11}$, 
(b) ${\rm Im}\, \hat{\Pi}^{00}$,
(c) ${\rm Im}\, \hat{\Pi}^{00}_{88}$, 
(d) ${\rm Im}\, \tilde{\Pi}^{00}$,
(e) $-{\rm Im}\, \tilde{\Pi}^{0i} \, \hat{p}_i$, 
(f) ${\rm Im}\, \tilde{\Pi}^\ell$,
(g) ${\rm Im}\, \Pi^t_{11}$, 
(h) ${\rm Im}\, \tilde{\Pi}^t$,
(i) ${\rm Im}\, \Pi^t_{88}$.
The corresponding results in the HDL limit, {\it i.e.\/} for $\phi=0$,
are shown as dotted lines.}
\label{fig1}
\end{figure}

\begin{figure}
\hspace*{2cm}
\mbox{\epsfig{figure=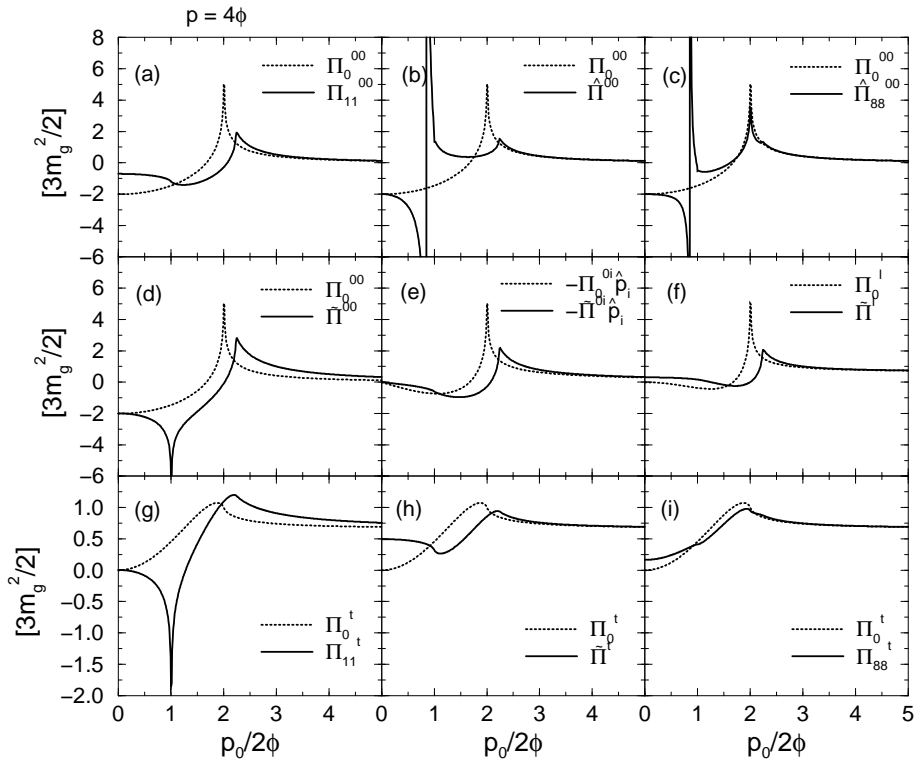,height=12cm,angle=270}}
\caption{The same as in Fig.\ \ref{fig1}, but for the
real parts.}
\label{fig2}
\end{figure}

At this point, we have determined all functions entering
the transverse and longitudinal polarization functions
for the eighth gluon. In Fig.\ \ref{fig1} we show the 
imaginary parts and in Fig.\ \ref{fig2} the real parts, for
a fixed gluon momentum $p= 4\, \phi$, as a function of
gluon energy $p_0$ (in units of $2\, \phi$). The units
for the imaginary parts are $-3 \, m_g^2/2$, and for
the real parts $+ 3\, m_g^2/2$.
For comparison, in parts (a) and (g) of these figures,
we show the results from Ref.\ \cite{dhrselfenergy} for
the longitudinal and transverse polarization function
of the gluon with adjoint color 1. In parts (d), (e), and (f)
the functions $\tilde{\Pi}^{00}$, $-\tilde{\Pi}^{0i}\, \hat{p}_i$,
and $\tilde{\Pi}^\ell$ are shown. 
According to Eq.\ (\ref{hatPi00}) these are required
to determine $\hat{\Pi}^{00}$, shown in part (b).
Using Eq.\ (\ref{Pi0088}), this result is then combined
with the HDL polarization function $\Pi^{00}_0$ to compute
$\hat{\Pi}^{00}_{88}$, shown in part (c).
Finally, the transverse polarization function for gluons of color 8
is shown in part (i). This function is given by the linear combination 
$\Pi^t_{88} = \frac{2}{3}\, \Pi^t_0 + \frac{1}{3}\, \tilde{\Pi}^t$
of the transverse HDL polarization function $\Pi^t_0$ and the
function $\tilde{\Pi}^t$, both of which are shown in part (h).
In all figures, the results for the two-flavor color superconductor
are drawn as solid lines, while the dotted lines correspond to
those in a normal conductor, $\phi \rightarrow 0$ (the HDL limit).

Note that parts (a), (d), (g), (h), and (i) of Figs.\ \ref{fig1}
and \ref{fig2} agree with parts (a), (b), (d), (e), and (f)
of Figs.\ 2 and 3 of Ref.\ \cite{dhrselfenergy}.
The new results are parts (e) and (f) of Figs.\ \ref{fig1} and \ref{fig2},
which are used to determine the functions in parts (b) and (c), the latter
showing the correct longitudinal polarization function for the
eighth gluon. In Ref.\ \cite{dhrselfenergy}, this function was not computed 
correctly, as the effect from the fluctuations of the condensate
on the polarization tensor of the gluons was not taken into account.

The singularity around a gluon energy somewhat smaller than
$p_0 = 2\, \phi$ visible in Figs.\ \ref{fig2} (b) and (c) seems
peculiar. It turns out that it arises due to a zero in the denominator of
$\hat{\Pi}^{00}$ in Eq.\ (\ref{hatPi00}), {\it i.e.}, when
$P_\mu\, \tilde{\Pi}^{\mu \nu}(P)\, P_\nu = 0$.
As discussed above, this condition defines the 
dispersion branch of the Nambu-Goldstone excitations \cite{zarembo}.
Therefore, the singularity is tied to the 
existence of the Nambu-Goldstone excitations of the diquark condensate.

\subsection{Spectral densities}

\begin{figure}
\hspace*{2cm}
\mbox{\epsfig{figure=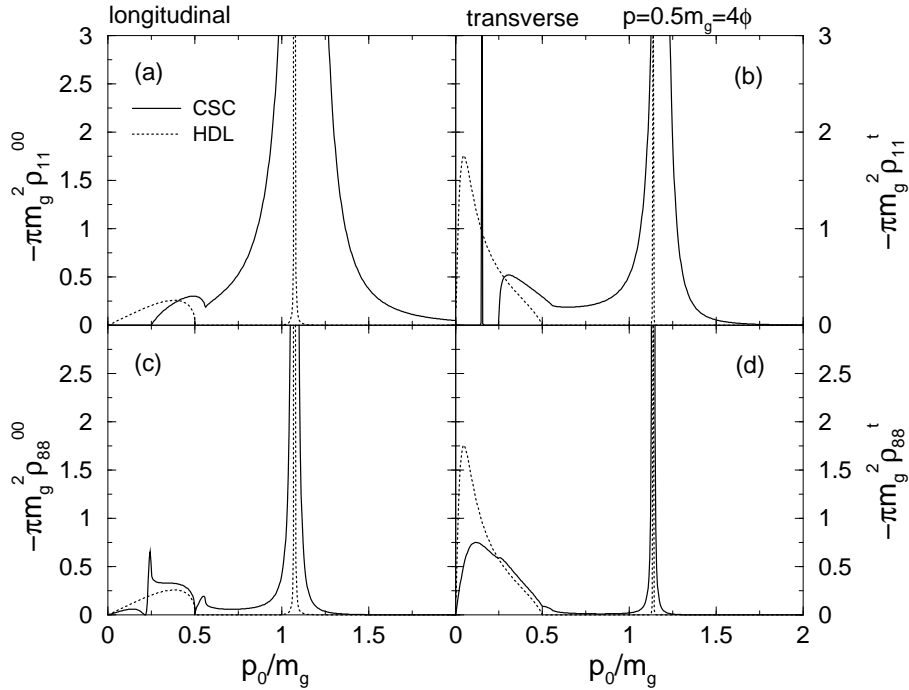,height=12cm,angle=270}}
\caption{The longitudinal (a), (c) and transverse (b), (d)
spectral densities for gluons of color 1 (a), (b) and
8 (c), (d). The gluon momentum is $p=m_g/2$ and $m_g = 8 \phi$.
For comparison, the dotted lines represent the
corresponding HDL spectral densities.
The poles of the spectral density corresponding to stable
quasiparticles are made visible by using
a numerically small but nonzero imaginary part.}
\label{fig3}
\end{figure}

Let us now determine the spectral densities for longitudinal
and transverse modes, defined by (cf.\ Eq.\ (45) of Ref.\ \cite{dhrselfenergy})
\begin{equation}
\rho^{00}_{88}(p_0, {\bf p}) \equiv \frac{1}{\pi}\,
{\rm Im}\, \hat{\Delta}^{00}_{88} (p_0 + i \eta, {\bf p}) 
\,\,\,\, , \,\,\,\,\,
\rho^t_{88}(p_0, {\bf p}) \equiv \frac{1}{\pi}\,
{\rm Im}\, \Delta^t_{88} (p_0 + i \eta, {\bf p}) 
\end{equation}
The longitudinal and transverse spectral densities for gluons
of color 8 are shown in Figs.\ \ref{fig3} (c) and (d), for fixed gluon momentum
$p = m_g/2$ and $m_g = 8\, \phi$. For comparison, the corresponding
spectral densities for gluons of color 1 are shown in parts (a) and (b).
Parts (a), (b), and (d) are identical to those of Fig.\ 6 of
Ref.\ \cite{dhrselfenergy}, part (c) is new and replaces Fig.\ 6 (c)
of Ref.\ \cite{dhrselfenergy}. 
One observes a peak in the spectral density around
$p_0 = m_g$. This peak corresponds to the ordinary longitudinal 
gluon mode (the plasmon) present in a dense (or hot) medium.

Note that the longitudinal spectral density for gluons of
color 8 vanishes at an energy 
somewhat smaller than $p_0 = m_g/4$. The reason is the
singularity of the real part of the gluon self-energy seen
in Figs.\ \ref{fig2} (b) and (c). The location of this point
is where $P_\mu \, \tilde{\Pi}^{\mu \nu}(P)\, P_\nu =0$, {\it i.e.},
on the dispersion branch of the Nambu-Goldstone excitations.

\begin{figure}
\vspace*{-1cm}
\hspace*{2cm}
\mbox{\epsfig{figure=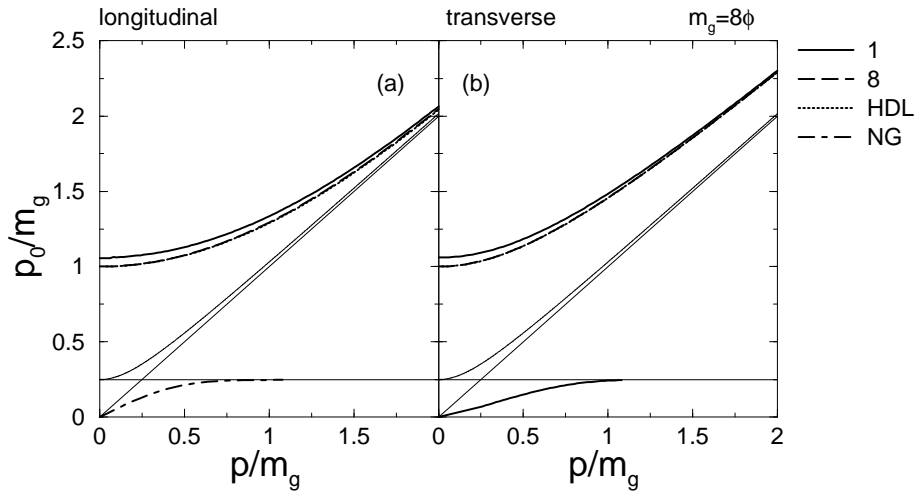,height=12cm,angle=270}}
\caption{Dispersion relations for (a) longitudinal and (b)
transverse modes for $m_g= 8\, \phi$.
The solid lines are for gluons of color 1, the dashed lines
for gluons of color 8. The dotted lines correspond to the
dispersion relations in the HDL limit.
For both longitudinal and transverse gluons of color 8, the
dispersion curves are indistinguishable from the HDL curves.
The additional branch shown in (a) as dashed-dotted line is the
one for the Nambu-Goldstone excitations, which appears as
a zero in the longitudinal spectral density.}
\label{fig4}
\end{figure}

Finally, we show in Fig.\ \ref{fig4} the dispersion relations for
all excitations, defined by the roots of 
\begin{mathletters} \label{disprel}
\begin{equation}
p^2 - {\rm Re}\, \hat{\Pi}^{00}_{88}(p_0, {\bf p}) = 0
\end{equation}
for longitudinal gluons (cf.\ Eq.\ (47a) of Ref.\ \cite{dhrselfenergy}),
and by the roots of
\begin{equation}
p_0^2 - p^2 - {\rm Re}\, \Pi^t_{88}(p_0, {\bf p}) = 0
\end{equation}
\end{mathletters}
for transverse gluons (cf.\ Eq.\ (47b) of Ref.\ \cite{dhrselfenergy}).
Let us mention that not all excitations found
via Eqs.\ (\ref{disprel}) correspond to
truly stable quasiparticles, {\it i.e.}, the
imaginary parts of the self-energies do not always vanish along
the dispersion curves. Nevertheless, in that case Eqs.\ (\ref{disprel})
can still be used to identify peaks in the spectral densities, which
correspond to {\em unstable\/} modes
(which decay with a rate proportional to the width of the peak).
As long as the width of the peak (the decay rate of the quasiparticles) 
is small compared to its height, it makes sense to refer to these
modes as quasiparticles.

Fig.\ \ref{fig4} corresponds to Fig.\ 5 of Ref.\ \cite{dhrselfenergy}.
In fact, part (b) is identical in both figures. 
Fig.\ \ref{fig4} (a) differs from Fig.\ 5 (a) of Ref.\ \cite{dhrselfenergy},
reflecting our new and correct results for the
longitudinal gluon self-energy. In Fig.\ 5 (a) of Ref.\ \cite{dhrselfenergy}, 
the dispersion curve for the longitudinal gluon of color 8 was
seen to diverge for small gluon momenta. In Ref.\ \cite{dhrselfenergy} 
it was argued that this behavior was due to neglecting 
the mesonic fluctuations of the diquark condensate.
Indeed, properly accounting for these modes, we obtain a
reasonable dispersion curve, approaching $p_0 = m_g$ as
the momentum goes to zero.
In Fig.\ \ref{fig4} (a) we also show the dispersion branch 
for the Nambu-Goldstone excitations (dash-dotted). 
This is strictly speaking not given
by a root of Eq.\ (\ref{disprel}),
but by the singularity of the real part of the
longitudinal gluon self-energy. However, because this singularity
involves a change of sign, a normal root-finding algorithm
applied to Eq.\ (\ref{disprel}) will also locate this singularity.
As expected \cite{zarembo}, the dispersion branch is linear, 
\begin{equation}
p_0 \simeq \frac{1}{\sqrt{3}}\, p \,\, ,
\end{equation}
for small gluon momenta, and approaches the value $p_0 = 2\, \phi$
for $p \rightarrow \infty$. 

\section{Conclusions} \label{IV}

In cold, dense quark matter with $N_f=2$ massless quark flavors,
condensation of quark Cooper pairs spontaneously breaks
the $SU(3)_c$ gauge symmetry to $SU(2)_c$. This results
in five Nambu-Goldstone excitations which mix with some
of the components of the gluon fields corresponding to the broken generators.
We have shown how to decouple them by a particular choice of 
't~Hooft gauge. The unphysical degrees of freedom
in the gluon propagator can be eliminated 
by fixing the 't~Hooft gauge parameter $\lambda = 0$.
In this way, we derived the propagator
for transverse and longitudinal gluon modes
in a two-flavor color superconductor
accounting for the effect of the Nambu-Goldstone excitations.

We then proceeded to explicitly compute the
spectral properties of transverse and longitudinal 
gluons of adjoint color 8. The spectral density of the longitudinal
mode now exhibits a well-behaved plasmon branch with
the correct low-momentum limit $p_0 \rightarrow m_g$.
Moreover, the spectral density vanishes for gluon energies and momenta
corresponding to the dispersion relation for Nambu-Goldstone excitations.
We have thus amended and
corrected previous results presented in Ref.\ \cite{dhrselfenergy}.

Our results pose one final question: using the correct expression
for the longitudinal self-energy of adjoint colors $4,\ldots,8$,
do the values of the Debye masses derived in Ref.\ \cite{dhr2f} change?
The answer is ``no''. In the limit $p_0 = 0,\, 
p \rightarrow 0$, application of Eqs.\ (120), (124), and (129)
of Ref.\ \cite{dhr2f} to Eq.\ (\ref{hatPi00aa}) yields
$\hat{\Pi}^{00}_{aa}(0) \equiv \Pi^{00}_{aa}(0)$, and
the results of Ref.\ \cite{dhr2f} for the Debye masses remain valid.

\section*{Acknowledgments}

We thank G.\ Carter, D.\ Diakonov, and R.D.\ Pisarski for discussions.
We thank R.D.\ Pisarski in particular for a critical
reading of the manuscript and
for the suggestion to use 't~Hooft gauge to decouple meson and
gluon modes. D.H.R.\ thanks the Nuclear Theory groups 
at BNL and Columbia University for their hospitality during a visit
where part of this work was done. He also gratefully acknowledges
continuing access to the computing facilities of
Columbia University's Nuclear Theory group.
I.A.S.\ would like to thank the members of the Institut f\"ur 
Theoretische Physik at the Johann Wolfgang Goethe-Universit\"at
for their hospitality, where part of this work was done.
The work of I.A.S.\ was supported by the U.S.\
Department of Energy Grant No.~DE-FG02-87ER40328.

\end{document}